\documentclass{aastex631}

\usepackage{comment}

\newcommand{\Fermi}{{\it Fermi}}
\newcommand{\fermi}{{\it Fermi}}

\newcommand{\lims}[2]{^{+#1}_{-#2}}

\newcommand{\revision}[1]{\textcolor{blue}{#1}}
\renewcommand{\revision}[1]{#1}

\usepackage{natbib}

\DeclareRobustCommand{\ion}[2]{%
\relax\ifmmode
\ifx\testbx\f@series
{\mathbf{#1\,\mathsc{#2}}}\else
{\mathrm{#1\,\mathsc{#2}}}\fi
\else\textup{#1\,{\mdseries\textsc{#2}}}%
\fi}

\usepackage{booktabs}
\usepackage{multirow}
\usepackage{amsmath}
\usepackage{array}
\newcolumntype{H}{>{\setbox0=\hbox\bgroup}c<{\egroup}@{}}
\shortauthors{McDaniel et al.}

\begin{document}

\title{Gamma-ray Emission from Galaxies Hosting Molecular Outflows}

\correspondingauthor{Alex McDaniel}
\email{armcdan@clemson.edu}

\author[0000-0002-8436-1254]{Alex McDaniel}
\affiliation{Department of Physics and Astronomy, 
Clemson University,
Clemson, SC, 29631}

\author[0000-0002-6584-1703]{Marco Ajello}
\affiliation{Department of Physics and Astronomy, 
Clemson University,
Clemson, SC, 29631}

\author[0000-0002-6774-3111]{Chris Karwin}
\affiliation{Department of Physics and Astronomy, 
Clemson University,
Clemson, SC, 29631}

\begin{abstract}
Many star-forming galaxies and those hosting active galactic nuclei (AGN) show evidence of massive outflows of material in a variety of phases including ionized, neutral atomic, and molecular. Molecular outflows in particular have been the focus of recent interest as they may be responsible for removing gas from the galaxy, thereby suppressing star formation. As material is ejected from the cores of galaxies, interactions of the outflowing material with the interstellar medium can accelerate cosmic rays and produce high-energy gamma rays. In this work, we search for gamma-ray emission from a sample of local galaxies known to host molecular outflows using data collected by the {\fermi} Large Area Telescope. We employ a stacking technique in order to search for and characterize the average gamma-ray emission properties of the sample. Gamma-ray emission is detected from the galaxies in our sample at the $4.4 \, \sigma$ level with a power-law photon index of $\Gamma \approx 2$ in the 1-800 GeV energy range. The emission is found to correlate with tracers of star formation activity, namely the $8-1000 \: \mu$m infrared luminosity. We also find that the observed signal can be predominantly attributed to \ion{H}{ii} galaxies hosting energy-driven outflows. While we do not find evidence suggesting that the outflows are accelerating charged particles directly, galaxies with molecular outflows may produce more gamma rays than galaxies without outflows. In particular, the set consisting of gamma-ray-detected galaxies with molecular outflows are nearly perfect calorimeters and may be future targets for searches of high-energy neutrinos.
\end{abstract}


\defcitealias{fluetsch19}{F19}
\defcitealias{ajello_sfgs}{A20}

\section{Introduction} \label{sec:intro}

The presence of galactic outflows and winds is well documented in galaxies over a wide range of distances and physical scales. Whether powered by starburst activity or active galactic nuclei, these winds are able to drive large amounts of material from their host galaxies, injecting energy into their surrounding medium \citep{veilleux2005, cicone2018nature, veilleux2020}. Galactic outflows manifest in a variety of different phases and with observational evidence spanning a wide range of frequencies. The sub-pc highly-ionized outflows are primarily measured by X-ray absorption lines \citep{reeves2009, tombesi2012, tombesi2013,gofford2013, nardini2015}, whereas the neutral atomic phase is primarily measured by observations of the sodium doublet \citep{heckman2000, rupke2005, cazzoli2016, robertsBorsani2019}, and the molecular phase is measured through various radio, infrared, and optical observations \citep{fischer2010,feruglio2010,sturm2011, combes13,spoon2013,veilleux2013, cicone14,garciaBurillo15, stone2016, GonzalezAlfonso2017, bolatto2021, stuber2021}. Together, understanding the details of the various phases of galactic outflows helps to shed light on {galactic} structure and feedback in galaxies.

Among the different outflow phases, the molecular phase is particularly interesting. For one, the molecular phase dominates the mass of the outflowing material and extends to the largest physical scales \citep{cicone14,carniani2015, garciaBurillo15}. Furthermore, the molecular gas driven in the wind is also the fuel for star formation, creating a direct link between the molecular outflow and star formation properties of the galaxy with potential effects on galaxy evolution. Detailed studies of molecular outflows have recently gained interest due in part to the capabilities of instruments such as {\it Herschel} in infrared (IR) or the Atacama Large Millimeter/submillimeter Array (ALMA) and the NOrthern Extended Millimeter Array (NOEMA) at millimeter wavelengths, which allow for several methods of detecting molecular outflows (see also \citealp{veilleux2020, stuber2021}). In infrared, P Cygni profiles of OH transitions have yielded multiple detections \citep{fischer2010,sturm2011, spoon2013,veilleux2013, stone2016, GonzalezAlfonso2017}, while CO line transitions (as well as other molecular tracers, e.g. HCN; \citealp{aalto2012}) with instruments such as ALMA, NOEMA, or the IRAM Plateau de Bure Interferometer (PdBI) have also provided an effective method for detecting molecular outflows and characterizing their properties \citep{feruglio2010,combes13, cicone14,garciaBurillo15,  bolatto2021}. Molecular outflows are the dominant component of the total outflow mass, have mass-loss rates on the order of a few hundred M$_{\odot}$ yr$^{-1}$, and extend to scales of $0.1-10$ kpc with wind velocities on the order of $10^2-10^3$ km s$^{-1}$ \citep{sturm2011, cicone14, fluetsch19,lutz2020, fluetsch2020}. They are found throughout the universe, from nearby systems out to as far as redshift of $z\sim 6$ \citep{jones2019,spilker2020}. Commonly – though not exclusively – they are found associated with (ultra)-luminous infrared galaxies ((U)LIRGs) \citep{puma2, chen2010, veilleux2013, perSant2018}. In a recent study by \citet{fluetsch19}, a collection of local ($z<0.2$) molecular outflows has been compiled from the literature and archival data in order to analyze their properties and examine the relations between them in a systematic manner.

While studies of galactic outflows have primarily been limited to the energy regimes of X-rays and below, theoretical models suggest that the interactions of the outflowing gas with the interstellar medium can create shocks in which cosmic rays can be accelerated. The cosmic rays can then interact with the ambient material and interstellar radiation fields to produce gamma rays through both hadronic and leptonic processes \citep{lamastra2016, wangLoeb}. The efficiency of cosmic-ray acceleration in outflows is predicted to be comparable to or in excess of other acceleration sites such as supernova remnants (SNRs, \citealp{FG2012, nims2015}). Recently, the detection of gamma rays from highly-ionized, ultra-fast outflows (UFOs) using \Fermi-LAT data has been reported \citep{karwinUFOs}, and it is possible that molecular outflows may also be observed in gamma rays \citep{lamastra2016}. In fact, several of the galaxies that are known to host powerful outflows are also gamma-ray emitters with significant detections by \Fermi-LAT \citep{lenain2010, Abdo_ngc253_m82, ackermann2012,circinus_gamma,  tang2014, ajello_sfgs}, as well as by other higher-energy gamma-ray telescopes \citep{acero2009_NGC253,veritas2009}. These include some notable and particularly well-studied systems, such as M 82, NGC 253, and NGC 1068. In some cases, the emission from gamma-ray-detected galaxies hosting molecular outflows exceeds that expected from $L_\gamma-L_{IR}$ relations \citep{ ajello_sfgs}.

Despite the theoretical basis for gamma-ray emission from molecular outflows and the gamma-ray detection of several galaxies hosting molecular outflows, thus far no concrete detection that can be directly attributed to the molecular outflow exists. Models of the gamma-ray emission from molecular outflows predict a relatively faint signal, which can be difficult to distinguish from other sources of gamma-ray emission such as starburst activity \citep{lamastra2016}. It is also unclear what the interplay may be between star formation activity and the molecular outflow. These two phenomena are intrinsically linked to the molecular gas of the galaxy \citep{feldmann2020} - in some cases, it has even been shown that enhanced star formation may take place within the outflow itself \citep{SFinOutflow, gallagher2019,arp220_SF_enhance}. 

The primary goal of this paper is to study the potential gamma-ray emission from a well-selected sample of galaxies that are known to host molecular outflows and that have not yet been individually resolved by current gamma-ray instruments. To do this we use $\sim 11$ years of {\fermi}-LAT data and employ a stacking technique designed to detect faint sources and characterize their emission. We then aim to determine the origin of the gamma-ray emission and how its properties relate to the properties of the molecular outflow and whether it can be disentangled from star-formation—induced gamma rays.

The remainder of the paper is as follows: in Section \ref{sec:sample}, we describe the sample of molecular outflows, while we describe the gamma-ray data selection and the analysis procedure in Section 3. In Section \ref{sec:results}, we present the results of the gamma-ray analysis and study the relationship between the gamma-ray emission and galaxy properties. In Section \ref{sec:conclusion}, we provide a discussion of these results. Throughout this work, we adopt cosmological values of $H_0 = 70$ km s$^{-1}$ Mpc$^{-1}$, $\Omega_M = 0.27$ and $\Omega_{\Lambda} = 0.73$.
\section{Sample Selection}\label{sec:sample}

The initial sample of molecular outflows in our analysis is taken from (\citet{fluetsch19}, hereafter \citetalias{fluetsch19}). In their work, they collect a sample of 45 galaxies with evidence of molecular outflows within the local universe ($z<0.2$). The sample includes 31 galaxies taken from the literature with outflow properties obtained through the analysis of the CO(1-0) and CO(2-1) emission lines using observations from either the IRAM PdBI \citep{cicone14,dasyra14,leroy15,garciaBurillo15,querejata} or ALMA \citep{combes13,sun14,pereiraSantaella,salak16,veilleux_obs}. An additional 10 outflows were identified from archival ALMA CO data by the authors of \citetalias{fluetsch19}. However, five of the 31 outflows taken from the literature and three of those from the ALMA archival data only include upper limits on the outflow properties. Also included in the sample of \citetalias{fluetsch19} are four outflows observed with far-infrared transitions of OH with the {\it Herschel}/PACS spectrometer  \citep{GonzalezAlfonso2017}. 

From this sample we remove a number of galaxies based on the following criteria: first, we remove the 8 non-detections wherein only upper limit estimates were provided by \citetalias{fluetsch19}, after which 37 outflows remain. We then proceed to make cuts based on spatial coincidence with sources in the {4FGL} \citep{4fgl} by removing all galaxies that fall within the 95\% confidence radius of a 4FGL source\footnote{The spatial coincidence cuts are performed using the first 4FGL data release \citep{4fgl} to be consistent with the point-source modeling. More recent 4FGL data releases do not detect additional galaxies from our benchmark sample, therefore the spatial coincidence cuts are the same for \revision{the subsequent 4FGL-DR2 \citep{DR2} and 4FGL-DR3 \citep{DR3} data releases.}}. This criterion removes 5 galaxies from the sample, each of which has been directly studied and detected by \Fermi. Specifically, these are NGC 2146 \citep{tang2014}, NGC 1068 \citep{ackermann2012}, the Circinus Galaxy \citep{circinus_gamma}, NGC 253, and M82 \citep{Abdo_ngc253_m82}. Although we exclude these from the stacking of unresolved sources, they are used in the later analysis (see Section \ref{sec:results}), and the {\fermi} data for these sources are analyzed following the same procedure described in Section \ref{sec:data_selection} that is applied to the benchmark sample. We additionally check for spatial coincidences with known gamma-ray blazars from the Roma-BZCAT catalog \citep{bzcat}  and remove bright radio galaxies included in the 3C/4C catalogs \citep{3c,4c} or in \citet{yuanWangcat}. For the BZCAT and radio sources, we remove any targets that lie within $0.1^{\circ}$ of the sources. This value is chosen as it is roughly similar to the mean 4FGL 95\% confidence radius. These criteria remove only one additional source -- the radio galaxy 4C 12.50. In all, the spatial coincidence cuts remove 6 galaxies from our sample. In addition, we identify two galaxies that have nearby, extremely bright 4FGL sources (although outside the 95\% confidence radius). Specifically, IRAS 05189-2524
 is $\sim 0.5^{\circ}$ away from 4FGL J0523.3-2527 (classified as a binary), and IRAS 15115+0208 is $\sim 0.45^{\circ}$ away from 4FGL J1512.2+0202, which is associated with the flat-spectrum radio quasar PKS 1509+022. The nearby 4FGL sources contribute relatively high counts and comprise the majority of the background within the 68\% containment radius of the point-spread-function (PSF) centered at the target. To avoid any potential impacts from these nearby, bright 4FGL sources, we remove the two targets from our sample. 

Our final benchmark sample consists of 29 galaxies. A selection of properties of the host galaxies and the molecular outflows are reported in Table \ref{tab:galaxy_params}. Several of these properties, such as optical classification, the AGN luminosity, and the AGN contribution to the bolometric luminosity ($\alpha_{bol} = L_{AGN}/L_{bol}$), are taken directly from \citetalias{fluetsch19} and references therein. Properties of the outflows such as mass-loss rates and kinetic power are derived from the line observations of the molecular outflows reported in \citetalias{fluetsch19}. To calculate the total 8 $\mu$m $-$ 1000 $\mu$m IR luminosity ($L_{\rm IR}$), we use the four IR flux bands ($f_{12\mu{\rm m}}$, $f_{25\mu{\rm m}}$, $f_{60\mu{\rm m}}$, $f_{100\mu{\rm m}}$) from the Infrared Astronomical Satellite (IRAS) Faint Source Catalog (FSC, \citealp{iras_fsc}) and the prescription of \citet{sanders1996}. Luminosity distances are taken from the NASA Extragalactic Database\footnote{\url{https://ned.ipac.caltech.edu/cgi-bin/objsearch?search_type=Search&refcode=2019MNRAS.483.4586F}} (NED).

To summarize, the general composition of our final sample includes 8 \ion{H}{ii} galaxies, 3 Seyfert 1 galaxies, 11 Seyfert 2 galaxies, and 7 LINERs (for simplicity, we will categorize all the LINERs and Seyferts as AGN galaxies throughout the remainder of the text, though it should be noted that $\alpha_{bol}$ is the more descriptive indicator of the role of the AGN contribution). The galaxies extend out to redshift $z \lesssim 0.2$, range in luminosity from $10^{9.5} L_{\odot} < L_{IR} < 10^{12.7} L_{\odot}$, and include 7 LIRGs and 15 ULIRGs.

\begin{table}[htbp]\label{tab:gal_params}
\centering

\begin{tabular}{lcHHcccccHHcccc}

\toprule
Name & Type & RA & Dec &$D_L$ & SFR & $\log L_{IR}$ & $\log L_{AGN}$ & $\alpha_{bol}$ & $R_{out}$ & $v_{out}$ & $\dot{M}_{out}$ & $\log P_k$ & $P_k/L_{AGN}$ &$q_{IR}$\\
 &  & [deg.]  & [deg.] &[Mpc] &  [$M_{\odot}$/yr] & [$L_{\odot}$] & [ergs/s] &  & [pc] & [km/s] & [$M_{\odot}$/yr] & [ergs/s]& & \\
\midrule
PG 0157+001 & Sy1 & 29.96 & 0.00 & 777.0 & 
            209.0 & 12.6 & 45.3 & 0.18 & 729.0 &268.0 &
            93.0 & 42.3 & $1.1 \times 10^{-3}$ & 2.13\\ 
NGC 1266 & LINER & 49.00 & -2.43 & 28.6 & 
            1.6 & 10.5 & 43.3 & 0.25 & 450.0 &177.0 &
            11.0 & 41.0 & $5.3 \times 10^{-3}$ & 2.36\\ 
IRAS F03158+4227 & Sy2 & 49.80 & 42.64 & 632.0 & 
            220.0 & 12.6 & 45.9 & 0.55 & 335.0 &1000.0 &
            1500.0 & 44.7 & 0.054 & --\\ 
NGC 1377 & LINER & 54.16 & -20.90 & 23.9 & 
            0.9 & 10.1 & 42.9 & 0.2 & 200.0 &110.0 &
            5.0 & 40.3 & $2.2 \times 10^{-3}$ & 3.53\\ 
NGC 1433 & Sy2 & 55.51 & -47.22 & 14.5 & 
            0.2 & 9.5 & 42.2 & 0.2 & 100.0 &100.0 &
            0.7 & 39.3 & $1.3 \times 10^{-3}$ & --\\ 
NGC 1614 & \ion{H}{ii} & 68.50 & -8.58 & 68.3 & 
            45.0 & 11.6 & $\leq$ 42.1 & $6.0 \times 10^{-4}$ & 560.0 &360.0 &
            21.0 & 41.9 & $\geq$ 0.731 & 2.77\\ 
NGC 1808 & \ion{H}{ii} & 76.93 & -37.51 & 10.8 & 
            5.1 & 10.7 & $\leq$ 41.0 & $5.0 \times 10^{-4}$ & 1000.0 &98.0 &
            3.0 & 40.0 & $\geq$ 0.095 & 2.82\\ 
IRAS F08572+3915 & Sy2 & 135.11 & 39.06 & 265.0 & 
            20.0 & 12.1 & 45.7 & 0.86 & 820.0 &800.0 &
            403.0 & 43.9 & 0.016 & 3.57\\ 
NGC 3256 & \ion{H}{ii} & 156.96 & -43.90 & 44.6 & 
            36.0 & 11.6 & $\leq$ 42.0 & $7.0 \times 10^{-4}$ & 500.0 &250.0 &
            4.0 & 40.9 & $\geq$ 0.085 & 2.37\\ 
IRAS F10565+2448 & Sy2 & 164.83 & 24.54 & 196.0 & 
            95.0 & 12.0 & 44.8 & 0.17 & 1100.0 &450.0 &
            100.0 & 42.8 & $9.9 \times 10^{-3}$ & 2.64\\ 
IRAS F11119+3257 & Sy1 & 168.66 & 32.69 & 929.0 & 
            144.0 & 12.7 & 46.2 & 0.689 & 7000.0 &1000.0 &
            203.0 & 43.8 & $4.0 \times 10^{-3}$ & 1.62\\ 
NGC 3628 & \ion{H}{ii} & 170.07 & 13.59 & 17.1 & 
            1.8 & 10.2 & $\leq$ 40.8 & $9.0 \times 10^{-4}$ & 400.0 &50.0 &
            1.5 & 39.1 & $\geq$ 0.019 & 2.00\\ 
ESO 320-G030 & \ion{H}{ii} & 178.30 & -39.13 & 51.1 & 
            20.0 & 11.1 & $\leq$ 41.1 & $1.0 \times 10^{-4}$ & 2500.0 &455.0 &
            1.2 & 40.9 & $\geq$ 0.637 & 2.77\\ 
NGC 4418 & Sy2 & 186.73 & 0.00 & 36.4 & 
            14.5 & 11.2 & 43.8 & $5.0 \times 10^{-4}$ & 569.0 &134.0 &
            19.0 & 41.0 & $1.7 \times 10^{-3}$ & 3.35\\ 
Mrk 231 & Sy1 & 194.06 & 56.87 & 189.0 & 
            234.0 & 12.5 & 45.7 & 0.34 & 600.0 &700.0 &
            350.0 & 43.7 & 0.01 & 2.44\\ 
IRAS 13120-5453 & Sy2 & 198.78 & -55.16 & 138.0 & 
            157.0 & 12.3 & 44.4 & 0.173 & 179.0 &549.0 &
            1115.0 & 44.0 & 0.474 & 2.78\\ 
M 51 & Sy2 & 202.48 & 47.23 & 11.1 & 
            2.6 & 10.4 & 43.8 & 0.61 & 37.0 &100.0 &
            11.0 & 40.5 & $5.6 \times 10^{-4}$ & 2.11\\ 
Mrk 273 & Sy2 & 206.18 & 55.89 & 169.0 & 
            139.0 & 12.1 & 44.2 & 0.08 & 550.0 &620.0 &
            200.0 & 43.4 & 0.168 & 2.49\\ 
SDSS J1356+1026 & Sy2 & 209.19 & 10.44 & 579.0 & 
            20.0 & 11.9 & 46.0 & 0.43 & 300.0 &500.0 &
            118.0 & 43.0 & $9.3 \times 10^{-4}$ & 2.32\\ 
IRAS F14348-1447 & LINER & 219.41 & -15.01 & 382.0 & 
            169.0 & 12.3 & 44.6 & 0.17 & 355.0 &450.0 &
            420.0 & 43.4 & 0.069 & --\\ 
IRAS F14378-3651 & LINER & 220.25 & -37.08 & 308.0 & 
            112.0 & 12.2 & 45.1 & 0.21 & 255.0 &425.0 &
            180.0 & 43.0 & $7.8 \times 10^{-3}$ & 2.27\\ 
NGC 6240 & Sy2 & 253.25 & 2.40 & 107.0 & 
            16.0 & 11.8 & 45.4 & 0.78 & 650.0 &400.0 &
            267.0 & 43.1 & $5.6 \times 10^{-3}$ & 2.10\\ 
IRAS 17208-0014 & \ion{H}{ii} & 260.84 & 0.00 & 189.0 & 
            200.0 & 12.4 & $\leq$ 43.7 & 0.24 & 160.0 &600.0 &
            176.0 & 43.3 & $\geq$ 0.427 & 2.79\\ 
NGC 6764 & LINER & 287.07 & 50.93 & 32.6 & 
            2.6 & 10.4 & 42.2 & 0.017 & 600.0 &170.0 &
            1.0 & 40.0 & $5.4 \times 10^{-3}$ & 2.25\\ 
IRAS 20100-4156 & \ion{H}{ii} & 303.37 & -41.79 & 605.0 & 
            330.0 & 12.7 & $\leq$ 42.9 & $7.0 \times 10^{-4}$ & 663.0 &456.0 &
            1457.0 & 44.0 & $\geq$ 11.23 & 2.93\\ 
IC 5063 & Sy2 & 313.01 & -57.07 & 47.2 & 
            0.6 & 10.8 & 44.3 & 0.9 & 500.0 &300.0 &
            8.0 & 41.4 & $1.1 \times 10^{-3}$ & 1.11\\ 
IRAS F20551-4250 & LINER & 314.61 & -42.65 & 187.0 & 
            43.0 & 12.0 & 44.8 & 0.13 & 175.0 &450.0 &
            200.0 & 43.1 & 0.023 & 2.89\\ 
IRAS 22491-1808 & \ion{H}{ii} & 342.95 & -17.87 & 348.0 & 
            145.0 & 12.1 & $\leq$ 41.6 & 0.06 & 202.0 &241.0 &
            654.0 & 43.1 & $\geq$ 27.454 & 3.26\\ 
IRAS 23365+3604 & LINER & 354.76 & 36.35 & 285.0 & 
            137.0 & 12.1 & 44.7 & 0.072 & 1230.0 &450.0 &
            57.0 & 42.6 & $7.8 \times 10^{-3}$ & 2.73\\ 

\bottomrule
\end{tabular}\caption{Galaxy and outflow properties for targets in the benchmark sample. For more detail see \citetalias{fluetsch19}. Luminosity distances are taken from NED, and the infrared luminosities ($L_{IR}$) are computed from the IRAS fluxes. The SFR is computed using $L_{IR}$, the AGN contribution to the total bolometric luminosity, and the relation of \cite{sturm2011}. The AGN contribution to the total bolometric luminosity is given by $\alpha_{bol}=L_{AGN}/L_{bol}$. $P_k$ is the kinetic power of the outflow, defined as $P_k=0.5\dot{M}_{out}v^2_{out}$. The values for $\alpha_{bol}$, $L_{AGN}$, mass loss rate, and outflow velocity are taken from \citetalias{fluetsch19}. $q_{IR}$ is the ratio of IR and 1.4 GHz radio fluxes, as defined in \citet{helou1985, ivison2010, harrison2014}, with radio fluxes taken from NED when available. Logarithmic values for $L_{IR}$, $L_{AGN}$, and $P_k$ are base 10 (i.e. $\log_{10}$).}\label{tab:galaxy_params}
\end{table}python check


\section{Data Selection and Analysis} \label{sec:analysis}
\subsection{Data} \label{sec:data_selection}
The data used in this analysis was collected over 11.1 years by the \Fermi-LAT between August 4, 2008 and September 10, 2019. We use events with energies in the range 1-800 GeV binned into 8 bins per decade and a pixel size of $0.08^{\circ}$. To reduce contamination from the Earth’s limb, we use a maximum zenith angle of $105^{\circ}$. We define a $10^{\circ}\times 10^{\circ}$ region of interest (ROI) centered at the position of each galaxy in the sample using RA and Dec values taken from NED. We use the standard data filters (DATA QUAL$>0$ and LAT CONFIG==1) and select photons corresponding to the P8R3\_SOURCE\_V2 class. The analysis is performed using \texttt{Fermipy} (v0.19.0, \citealp{fermipy}), which utilizes the underlying Fermitools (v1.2.23). The Galactic diffuse emission is modeled using the standard interstellar emission model (\texttt{gll\_iem\_v07.fits}). For the extragalactic emission and residual instrumental background we use \texttt{iso\_P8R3\_SOURCE\_V2\_v1.txt}, and the point source emission is modeled using the 4FGL catalog (\texttt{gll\_psc\_v20.fits}). In order to account for photon leakage from sources outside of the ROI due to the PSF of the detector, the model includes all 4FGL sources within a $15^{\circ}\times 15^{\circ}$ region. The energy dispersion correction (edisp\_bins=-1) is enabled for all sources except the isotropic component.

\subsection{Stacking Analysis}\label{sec:stacking_analysis}
\begin{figure}
    \centering
    \includegraphics[width=3.5in]{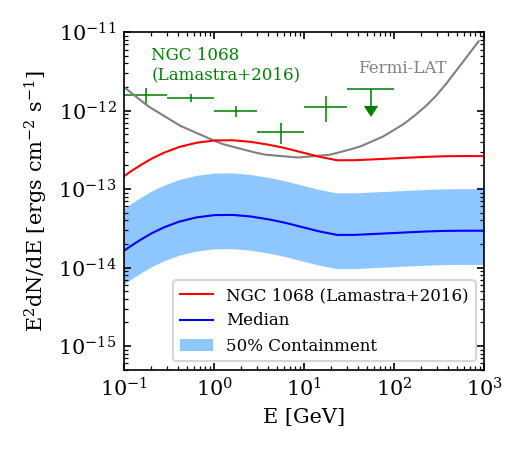}
    \caption{Gamma-ray SED for the molecular outflow model of NGC 1068 from \citet{lamastra2016} (red). The blue line and band show the predicted median and 50\% containment band of our sample when applying the same model scaled to the characteristics of each source. \revision{Data points for NGC 1068 from the \citet{lamastra2016} analysis are shown as green crosses.} We also show in grey the \Fermi-LAT broadband sensitivity for a power-law source using 10 years of {\Fermi} data.}
    \label{fig:MO_Model}
\end{figure}

While a number of galaxies hosting molecular outflows have also been observed in gamma rays by the \Fermi-LAT, it is not currently known to what extent the molecular outflow contributes to this emission. In most galaxies, it is likely that any potential molecular-outflow induced gamma-ray emission would fall under the detection threshold of the {\fermi}-LAT. To illustrate this, we consider the analysis of \cite{lamastra2016} wherein the gamma-ray emission of an AGN-driven molecular outflow is estimated for the gamma-ray detected galaxy NGC 1068 -- a relatively close ($D_L=14.4$ Mpc), bright, and particularly well-studied galaxy. Based on models that assume typical parameter values for $L_{AGN}$ and outflow characteristics of NGC 1068, as well as adopting conventions of SNR shock efficiencies for energy injection to cosmic ray protons and electrons, their results yield flux values at the level of roughly $\sim 2-5 \times 10^{-13}$ ergs cm$^{-2}$ s$^{-1}$ in the 1-800 GeV energy range. Making use of the \cite{lamastra2016} results and adopting their model wherein the gamma-ray emission from an outflow is directly related to the kinetic power of the outflow ($P_k = 0.5 \dot{M}_{out}v_{out}^2$), we can produce a rough estimate for gamma-ray emission from our sample. For each outflow, we scale the gamma-ray luminosity found in \cite{lamastra2016} by the kinetic power of the outflow, then calculate the flux from using the distance to the outflow. This gives a median expected flux on the order of $\sim 2-4 \times 10^{-14}$ ergs cm$^{-2}$ s$^{-1}$ (corresponding to a photon flux of $2.8 - 5.6 \times 10^{-12}$ ph cm$^{-2}$ s$^{-1}$ for a power law index of 2.2). This is illustrated in Figure \ref{fig:MO_Model}.  We emphasize that the \cite{lamastra2016} model represents the predicted contribution from only the molecular outflow. Their work finds that this is not sufficient to fully account for the gamma-ray detection of NGC 1068 (see Figure \ref{fig:MO_Model}), assuming the standard cosmic-ray acceleration efficiency parameters. Rather, a comparable contribution from starburst activity would be required to account for the gamma-ray emission. For comparison, we also show the \Fermi-LAT broadband sensitivity\footnote{\url{https://www.slac.stanford.edu/exp/glast/groups/canda/lat_Performance.htm}} for a point source located at intermediate latitudes ($\ell=0^{\circ}, \: b=30^{\circ}$) using 10 years of \Fermi-LAT data.

{The above estimates serve as an indication that the emission from individual molecular outflows is likely below the sensitivity of the {\Fermi}-LAT, and therefore motivates the use of a stacking technique} in order to detect emission from the overall population. The method employed is the same as that applied successfully in a number of previous studies (e.g. \citealp{EBL_stacking, Paliya2019, ajello_sfgs, karwinUFOs}). For this procedure, we work under the assumption that the sample population can be characterized by average quantities such as flux, luminosity, or photon index. We begin the analysis by optimizing the model components for the ROI of each target using a maximum likelihood fit and evaluate the significance of each source in the ROI using the TS defined by:
\begin{equation}\label{eq:TS}
     \mathrm{TS} = -2\log(\mathcal{L}_0/\mathcal{L}),
\end{equation}
 \noindent where $\mathcal{L}_0$ is the likelihood for the null hypothesis {(i.e. all sources except for molecular outflow), and $\mathcal{L}$ is the likelihood for the alternative hypothesis (all sources including the molecular outflow).}
Here, the spectral parameters of the Galactic diffuse component (index and normalization) and the normalization of the isotropic component are left free. We also leave free the normalizations of all 4FGL sources with TS $\geq$ 25 that are within $5^\circ$ of the ROI center, as well as sources with TS $\geq$ 500 and within $7^\circ$. The fitting of the molecular outflow source assumes a power-law spectral model with the normalization and index left free. At this stage, we also use the Fermipy function \textit{find\_sources} to search for new point sources. The \textit{find\_sources} function generates TS maps and identifies new sources based on peaks in the TS. The maps are generated using a power-law spectral model ($\frac{dN}{dE}\propto E^{-\Gamma}$) with an index of $\Gamma=2.0$. The minimum separation between two point sources is set to $0.5^\circ$, and the minimum TS for including a source in the model is set to 16. 

After these processing steps, we then create a bi-dimensional TS array in flux-index space for each target. The flux-index stacking method employed here has been validated a number of previous times through simulations (see e.g. \citealp{Paliya2019, ajello_sfgs, karwinUFOs}), and has been shown to be a reliable technique. Underpinning this approach is the assumption that if the gamma-ray emission in each target comes from the same emission mechanism, the average index will be broadly representative of the population. Similarly, the flux of the sample population is assumed to be roughly concentrated around the average, which is motivated by the fact that most {\fermi} sources are detected in flux near the threshold \citep{4fgl}. Furthermore, sources with particularly high fluxes are more likely to have already been individually detected. Some other stacking analysis studies have chosen to instead test alternative hypotheses, such as for example stacking one dimensional TS profiles as a function of only index \citep{deMenezes_stars}. However, we elect to follow the approach of generating the two dimensional TS profiles, which has been both successfully employed in previous studies as well as validated through several simulations.

 With the isotropic and galactic diffuse background models left free, we scan photon indices from 1 to 3.3 with a spacing of 0.1 and total integrated photon fluxes from $10^{-13}$ to $10^{-9}$ $\mathrm{ph \ cm^{-2}\ s^{-1}}$ with 40 logarithmically spaced bins {over the 1-800 GeV energy range. This choice of energy range is consistent with that used in the most recent application of this stacking analysis studying ultra-fast outflows \citep{karwinUFOs}.} Since the TS is an additive quantity, the stacked profile is merely the sum of the arrays for either the given sample or any desired sub-sample.


\section{Results}\label{sec:results}
\subsection{Stacked TS for the Benchmark Sample}
\begin{figure}
    \centering
    \includegraphics[width=3.5in]{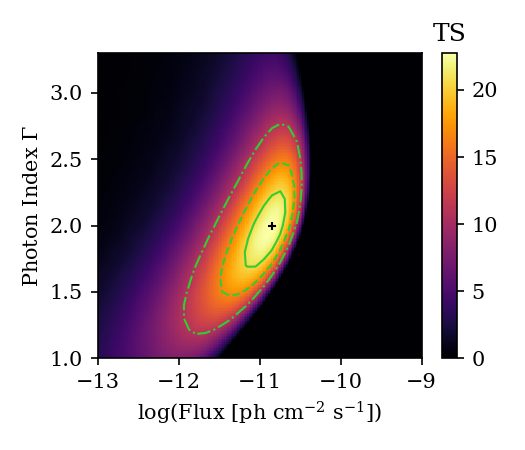}
    \caption{Stacked TS profile for the benchmark sample. Overlaid are the 1, 2, and $3\, \sigma$ contours {for 2 degrees of freedom}.}
    \label{fig:TS_benchmark}
\end{figure}

In Figure \ref{fig:TS_benchmark}, we show the stacked TS array for the full benchmark sample of 29 molecular outflows that have not previously been detected by gamma-ray observations. The best-fit photon flux is $1.3\lims{0.7}{0.6}\times 10^{-11}$ ph cm$^{-2}$ s$^{-1}$ with photon index $\Gamma=2.0\lims{0.3}{0.2}$. 
The maximal TS value is 22.8, corresponding to roughly a $4.4\,\sigma$ detection for 2 degrees of freedom. From the benchmark sample, we check for any galaxies in the sample that may be individually detected at a significant level. We note that none of the individual targets have significant (TS $>$ 25) detections; furthermore, all are below the $3\sigma$ level with a median TS value at the best-fit parameters of 2.2.

\subsection{{Scientific} Control Sample}\label{sec:alma_control}
\begin{figure}[tbph!]
    \centering
    \includegraphics[width=3.5in]{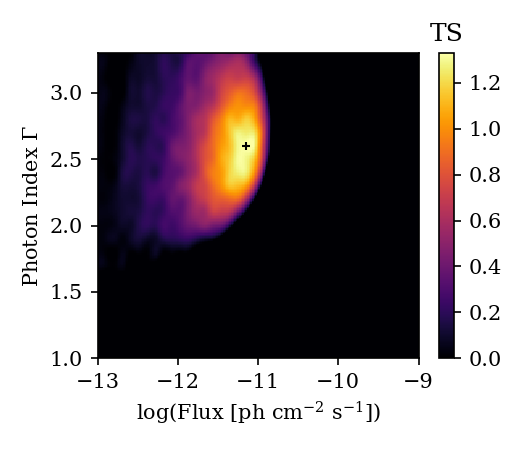}\hfill
    \includegraphics[width=3.5in]{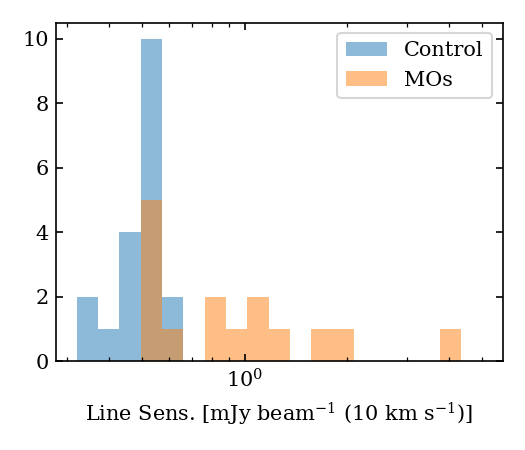}
    
    \caption{Left: TS profile for control sample galaxies where no outflow has been successfully detected. Right: Distribution of estimated line sensitivities from the ALMA archive for the best observations of the molecular outflows in our benchmark sample (orange) and those used in the scientific control sample (blue).}
    \label{fig:Control}
\end{figure}

In order to understand to what extent the purported signal from the benchmark sample can be attributed to the presence of the molecular outflows, we repeat our analysis on a control sample consisting of galaxies where no molecular outflow has been detected. In compiling the sample of molecular outflows in \citetalias{fluetsch19}, the authors analyzed ALMA archival data for $\sim 100$ galaxies in order to search for evidence of outflows. As discussed earlier (cf. Section \ref{sec:sample}), the authors were able to detect or even constrain outflow properties in only 10 galaxies. For our control sample, we therefore make use of a sub-sample of the galaxies for which no outflow was detected. However, it is important to note that most of these galaxies lack ALMA observations that are sensitive enough to detect the outflows, and non-detections do not necessarily imply the absence of outflowing molecular gas. We thus use the subset of galaxies with the most sensitive ALMA observations that were examined by the authors of \citetalias{fluetsch19}. The distribution of ALMA line sensitivities as presented in the ALMA Science Archive\footnote{\url{https://almascience.nrao.edu/aq/}}
is shown in Figure \ref{fig:Control} along with the most sensitive observations for the galaxies in the benchmark sample that were detected with ALMA observations. We note that all the galaxies selected for our control sample of ALMA non-detections have estimated sensitivities less than $\sim 0.62$ mJy beam$^{-1}$, better than for most of the detected galaxies. We therefore treat this as a reasonable selection of galaxies lacking a prominent molecular outflow. 

In constructing the control sample, our aim is to match the characteristics of the benchmark sample, particularly their distributions in distance and $L_{IR}$. However, the control sample obtained from the ALMA archival data poorly samples higher IR luminosities. For instance, only one galaxy from the ALMA archival control sample (IRAS 07251-0248) has an IR luminosity greater than $10^{12} L_{\odot}$, whereas almost half our benchmark sample has IR luminosities above this level. To address this, we searched the literature for known (U)LIRGs (i.e. $L_{\rm IR} \gtrsim 10^{12} L_{\odot}$) for which a search for a molecular outflow has been performed. We found no evidence of molecular outflows for these galaxies reported in the literature.

Previous studies -- particularly ones interested in the multi-phase nature of outflows -- have similarly searched the literature for evidence of the presence of various outflows in galaxies. Such searches have identified several candidates that lack any significant evidence of molecular outflows using a variety of detection techniques. Of these, we select IRAS 06259-4708N, IRAS 13156+0435N, and IRAS 19542+1110 \citep{fluetsch2020},  as well as IRAS 06035-7102, IRAS 00198-7926, and IRAS 20414-1651 \citep{westmoquette2012}. Additionally, in \citet{veilleux2013}, non-detections of an outflow in the molecular phase using {\it Herschel}/PACS observations of the OH 119 $\mu$m line were reported for PG 2130+099, IRAS F23128-5919, IRAS F15206+3342, IRAS F13305-1739. Finally, we also include the galaxy I Zw 1, which has been observed to have outflows in the neutral atomic and ionized phases, though the molecular outflow phase has not been directly constrained.  I Zw 1 was reported as a non-detection using CO emission lines in \citet{cicone14} and is listed as lacking evidence of a molecular outflow in \citet{fluetsch2020} (though the properties of the other phases were used to place limits on the molecular outflow in \citetalias{fluetsch19}). A search for more recent studies of the presence of molecular outflows in the systems listed above yields no definitive evidence.

We note that while these galaxies form one of the best control samples of ULIRGs lacking direct evidence of molecular outflows available, it is not necessarily the case that the presence of outflowing molecular gas can be explicitly excluded. In fact, when searching through catalogs of local known (U)LIRGs (e.g. in the IRAS Revised Bright Galaxy Survey (RBGS)  or Great Observatories All-sky LIRG Survey (GOALS) catalogs \citep{ rbgs,goals}), most candidates that have been studied tend to show some evidence of a molecular outflow. Furthermore, although the exact prevalence is not known, there is increasing evidence that molecular outflows are widely ubiquitous in these systems \citep{ chen2010, veilleux2013, perSant2018, puma2}. Therefore, we caution that this subset of the control sample should be thought of as a collection of galaxies where the molecular outflow is not prominent enough to be detected with standard techniques, rather than claiming that they are concretely excluded.

The control test analysis is performed following an identical procedure to the benchmark sample, including all spatial coincidence cuts discussed in Section \ref{sec:sample}. In total, this sample comprises 19 from the ALMA archival observations and 11 (U)LIRGS from the literature. The TS array for the control is shown in the left panel of Figure \ref{fig:Control}, with a peak value of TS = 1.3 at index of $\Gamma=2.6$ and  95\% upper limit of $1.6\times 10^{-11}$ ph cm$^{-2}$ s$^{-1}$. Given the relatively low TS for the control sample in comparison with the results for the benchmark sample, the conclusion that the observed signal is related to the presence of the outflow is supported.

\subsection{Technical Control Sample}\label{sec:technical_control}
\begin{figure}[!htbp]
    \centering
    \includegraphics[width=3.5in]{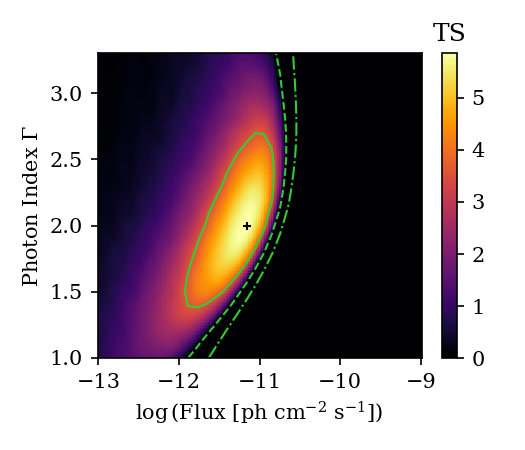}\hfill
    \includegraphics[width=3.5in]{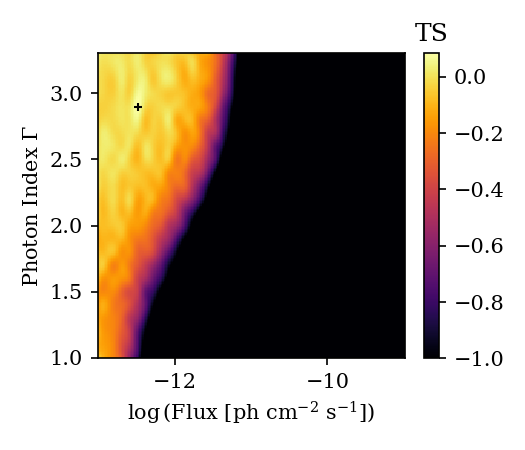}
    \caption{TS profiles for the iterations of the technical control sample with the highest maximum TS value (left) and the lowest maximum TS value (right). Note that for visualization purposes the minimum of the color scale for the right panel is set to $-1$.}
    \label{fig:technical_control_TS}
\end{figure}

As an additional test to the scientific control sample using galaxies without detected molecular outflows, we run a separate technical control analysis to account for systematic effects of the {\Fermi} analysis, such as the underlying background intensity and the effects of nearby gamma-ray sources in our model. This is performed in the following manner. For each galaxy in our benchmark sample of galaxies hosting molecular outflows (i.e. Table \ref{tab:galaxy_params}), we randomly select a set of coordinates located between $1^{\circ}-2^{\circ}$ from the galaxy coordinates and run these through the analysis pipeline. As in the previous cases, a bi-dimensional TS profile is created for each set of coordinates. We then stack the TS profiles to obtain an estimate of the background TS. This process is repeated five times, yielding in maximum TS values of TS $= [2.04, 5.86, 3.67, 0.86, 0.08]$.
Figure \ref{fig:technical_control_TS} shows the TS profiles for the iterations yielding the highest and lowest TS values. Although the fluctuations in the technical control TS can vary as high as TS $\approx 6$, the generally low TS values found in the control analysis indicate that our ROIs are well modeled.

\subsection{Radio Emission and the Role of Jets}\label{radio}
 Another potential contribution to the gamma-ray emission may be from jets. Particularly,  the presence of radio jets has been shown to be important for detecting a gamma-ray counterpart, e.g. in radio galaxies \citep{abdo2010} and low-luminosity AGN \citep{menezes2020}. One way to infer the presence of a radio jet is based on the ratio of the $8-1000\:\mu$m IR flux to the 1.4 GHz monochromatic radio
flux \citep{ivison2010, harrison2014}. Specifically, ratio values of $q_{IR}\lesssim 1.8$ are indicative of a radio excess and the likely presence of a radio jet, whereas values of $\sim 2.4$ are consistent with radio emission due to star-formation \citep{helou1985,ivison2010, harrison2014}. In Table \ref{tab:galaxy_params},  the $q_{IR}$ values for our sample are listed (see also Figure 13 of \citetalias{fluetsch19}). Two of the galaxies in the sample exhibit a radio excess, however for the majority of the sample there is little evidence for the presence of radio jets as indicated by these values. Given the relationship between radio and gamma-ray jets and the lack of evidence for jets in our sample, we expect any gamma-ray contributions from jet emission to be minimal. Furthermore, the lack of radio excess in these targets may indicate that the outflows are not accelerating large amounts of cosmic rays as cosmic-ray electrons would produce radio emission.


\subsection{Gamma Rays in Energy or Momentum Conserving Outflows}\label{sec:P_L_split}
The mechanisms by which the outflows are driven {in AGN galaxies} can be summarized by three theoretical paradigms. Two of the paradigms are directly related to the dynamics of the shock blast. In the energy-driven case, the shock expands in an adiabatic, energy conserving fashion due to inefficient cooling. In the momentum-driven case, the cooling of the shocked gas is more efficient, and the full energy of the wind is injected into the ambient medium. These two models are also often referred to as “energy-conserving” and “momentum-conserving,” respectively \citep{king2010, king2011,FG2012,kingPounds2015}. A third class of models for driving the outflow invokes the radiation-pressure—driven scenario in which the outflows can be driven by the direct pressure of IR, UV, and optical photons on the ISM \citep{fabian2012,ishibashi2018}. For star-formation-driven outflows, the canonical paradigm is an energy-driven scenario \citep{ chevalierClegg1985, heckman1990, cicone2016}. An alternate scenario where radiation pressure drives the outflow may also play a meaningful role \citep{murray2005,thompson2015}. However, for this to be the primary driver, a ratio of outflow momentum rate to the radiation momentum 
($\frac{\dot{M}_{out}v_{out}}{L_{bol}/c}$) near unity would be expected \citep{murray2005, davies2019}, whereas for most of the star-forming galaxies this ratio is $\sim 0.1-0.5$ \citepalias{fluetsch19}.

A useful distinction between the various AGN outflow models is the relation between the kinetic properties of the outflow and their host AGN. Specifically, the ratio of kinetic power ($P_k=0.5 \dot{M}_{out}v^2_{out}$) to the AGN luminosity $L_{AGN}$ in the energy-driven scenario is around 0.05 or greater \citep{ king2011,costa2014,kingPounds2015}. However, in momentum-driven models, the wind is less efficient at removing material from the inner regions of the galaxy, and a lower fraction of the AGN luminosity is transferred to the outflow, with $P_k/L_{AGN}$ values typically below $\sim ~0.1\%$ (\citealp{costa2014,kingPounds2015}; \citetalias{fluetsch19}). Additionally, radiation-pressure models show power fractions up to $\sim 1 \%$ or even superlinear scaling between the kinetic power and AGN luminosity \citep{ishibashi2018}. Typically, models favor the energy-driven mechanism for observations of large scale outflows, largely because the influence of the momentum-driven outflows is expected to be confined to the inner 0.1-1 kpc regime \citep{king2011, kingPounds2015}.

\begin{figure}
    \centering
    \includegraphics[width=3.5in]{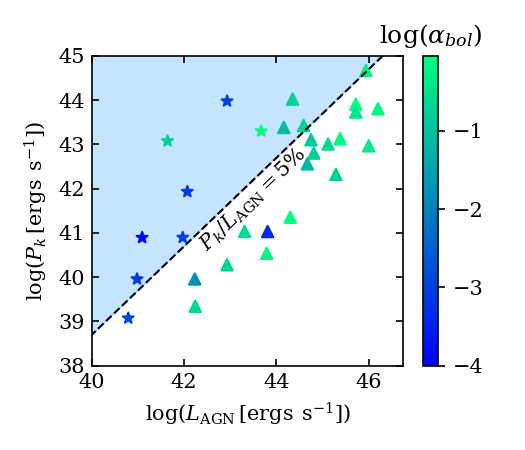}\hfill
    \includegraphics[width=3.5in]{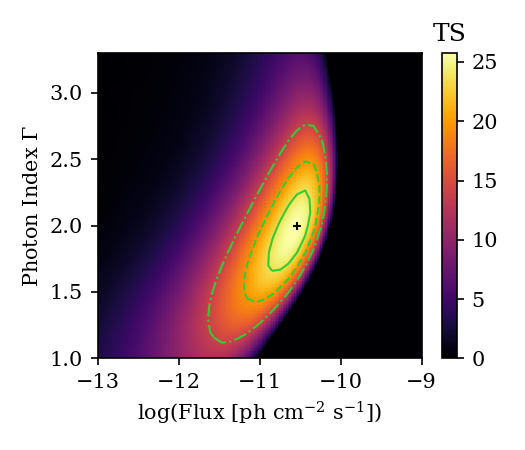}
    \caption{Left: Distribution of the benchmark sample in the $P_k - L_{AGN}$ plane. The shaded area is the region corresponding to the energy-driven regime above the $P_k/L_{AGN}=5\%$ line. The shaded region contains 11 of the outflows, while the remaining 18 fall into the $P_{k}/L_{AGN}<5\%$ region. Points with star markers are the \ion{H}{ii} galaxies, while the triangles represent AGN galaxies. Colors correspond to the log of the AGN contribution to the bolometric luminosity ($\alpha_{bol} = L_{AGN}/L_{bol}$). Right: Stacked TS array for the galaxies in the energy-driven (i.e.  $P_k/L_{AGN}>5\%$) regime. The max TS is 25.8 and the best-fit flux and index are $2.5\lims{1.4}{1.3}\times 10^{-11}$ ph cm$^{-2}$ s$^{-1}$ and $\Gamma=2.0\lims{0.3}{0.3}$, respectively.} \label{fig:P_L_split}
\end{figure}

Here, we explore whether there is any connection between the adopted driving mechanism and the observed gamma-ray emission. To do this, we separate the galaxies in our sample {at the 5\% value for the ratio of the outflow kinetic power to the AGN luminosity, roughly grouping the sample into galaxies that fall into the energy-driven regime from those that do not (i.e. they are more consistent with momentum-driven or radiation-pressure models)}. In the left panel of Figure \ref{fig:P_L_split}, we show the distribution of our sample in the $P_k-L_{AGN}$ space along with the 5\% line. In the right panel of Figure \ref{fig:P_L_split}, we show the stacked TS profile for the subsample of energy-driven outflows, which yields a max TS of 25.8 at a flux and index of $2.5\lims{1.4}{1.3}\times 10^{-11}$ ph cm$^{-2}$ s$^{-1}$ and $\Gamma=2.0\lims{0.3}{0.3}$, respectively. On the other hand, in the non-energy-driven regime, the maximum TS value is only TS = 3.6. Thus, we see that any signal from our sample coincides with the energy-driven subset and even slightly improves upon the signal of the full benchmark sample.

Since the production of gamma rays from molecular outflows relies on cosmic ray interactions in the ISM, it appears consistent that the signal would most coincide with energy-driven outflows. In this paradigm, the outward expanding gas propagates more quickly and imparts greater momentum to the ISM, in comparison to momentum-driven outflows where most of the energy is lost in cooling processes within $\sim 1$ kpc scales from the launched winds.

{As can be seen in Figure \ref{fig:P_L_split}, the subsets created by the $P_k/L_{AGN}=5\%$ division also subdivide the sample into groups containing either mostly  \ion{H}{ii} or mostly AGN galaxies. For comparison, we compute the signal from the explicit subgrouping based on  \ion{H}{ii} or AGN classification of the galaxy. We find that the subset of only \ion{H}{ii} galaxies has TS = 22.45, while the subset of only AGN galaxies have TS = 6.65.}

\subsection{Gamma-ray Luminosity Scaling Relations}
In the following section, we investigate the scaling relationship between the gamma-ray luminosity and the properties of the outflow sample. As a recent example, \citet{karwinUFOs} found that the gamma-ray emission from ultra-fast outflows scales with both the bolometric luminosity of the host and the kinetic power of the outflow. In star-forming galaxies, strong correlations exist with radio and IR luminosities \citep{ackermann2012, ajello_sfgs}. For our approach, we assume a simple log-linear relation of the form:
\begin{equation}\label{eq:Lgamma_gen}
    \log_{10} \left(\frac{L_{\gamma}}{\mathrm{ergs/s}}\right) = \beta + \alpha \log_{10} \left(\frac{X}{X_0}\right),
\end{equation}
where $X$ is some parameter of interest normalized to $X_0$. {For each target, we convert the flux-index plane to $\alpha-\beta$ space using the known distance and adopting the best-fit photon index found in the stacked flux-index TS profile (i.e. $\Gamma=2$, see Figure \ref{fig:TS_benchmark}). We then combine the individual $\alpha-\beta$ TS profiles to obtain the stacked TS profile in the $\alpha-\beta$ plane.} We investigated possible trends with a number of different properties of the host galaxy and the outflow itself. \citetalias{fluetsch19} provides several characteristics of the host galaxies and the outflows, either aggregated from the literature or (in the case of the outflow parameters in their archival ALMA outflows) calculated from the data directly. We explored several parameters of interest using the relation above (particularly, the AGN bolometric luminosity, the outflow kinetic power, and the mass outflow rate); however, in most cases the relation found was not significant and/or provided a lower TS than the simple flux stacking shown in Figure \ref{fig:TS_benchmark}. Of the parameters considered, the IR luminosity provided the strongest correlation {and the only one showing improvement over the benchmark flux-index TS of 22.8 (with the exception of the AGN corrected SFR, see the end of section \ref{sec:SFG}).}

\subsubsection{The $L_{\gamma} - L_{IR}$ Correlation}
\begin{figure}
    \centering
    \includegraphics[width=3.5in]{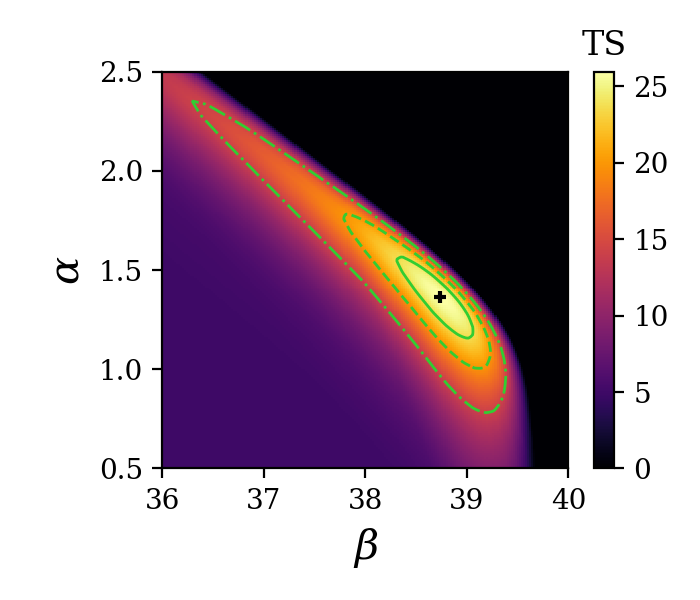} 
    \hfill
    \includegraphics[width=3.5in]{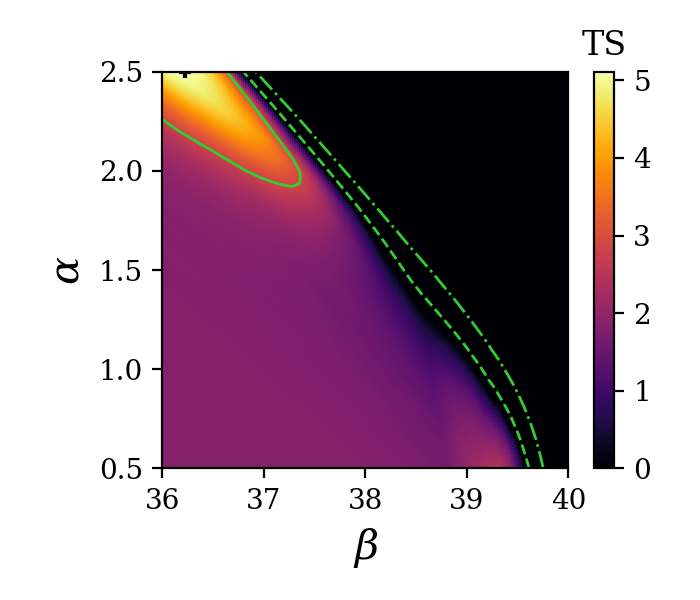}
    \caption{TS profiles in the $\alpha$—$\beta$ plane, characterizing the $L_{\gamma}$—$L_{IR}$ relation (see Equation \ref{eq:Lgamma_LIR}) for the benchmark sample {(left panel) and for the scientific control sample (right panel). Contours show the 1, 2, and 3 $\sigma$ levels for 2 degrees of freedom.}}
    \label{fig:TS_alpha_beta}
\end{figure}

From the various relations explored, the strongest correlation found was between the gamma-ray emission and the infrared luminosity, stated explicitly as:
\begin{equation}
    \log_{10} \left(\frac{L_{\gamma}}{\mathrm{ergs/s}}\right) = \beta + \alpha \log_{10} \left(\frac{L_{\mathrm{IR}}}{10^{10}L_{\odot}}\right).\label{eq:Lgamma_LIR}
\end{equation} 
For this relation, we find best fit values of $\alpha = 1.36\lims{0.08}{0.12}$ and $\beta = 38.7\lims{0.16}{0.20}$ with a TS = 25.9, a $\Delta TS = 3.1 $ improvement over the flux-index stacking. This suggests that there is a significant relationship between the gamma-ray and infrared luminosities for this sample. The resulting TS profiles in the $\alpha$—$\beta$ space for the benchmark and scientific control samples are shown in the left and right panels of Figure \ref{fig:TS_alpha_beta}. Numerous previous studies have established the connection between a galaxy’s infrared and gamma-ray luminosities \citep{ackermann2012, ajello_sfgs}. The standard interpretation for this is that both emission types can be traced to star-formation activity. The infrared is a result of the UV light of massive stars being absorbed and re-emitted by the interstellar dust \citep{lonsdaleHelou1987,buatXu1996}, whereas the gamma rays are produced from cosmic rays accelerated by the core collapse supernovae of massive stars. A number of star-forming galaxies have been directly detected at gamma-ray energies by {\Fermi}-LAT (see e.g. \citealp{ackermann2012, ajello_sfgs, kornecki2020}); however, there are many more that have yet to be detected. In  \citet[][hereafter \citetalias{ajello_sfgs}]{ajello_sfgs}, a stacking analysis similar to the one performed here was conducted on a sample of star-forming galaxies with the goal of characterizing the gamma-ray emission in both detected galaxies and undetected star-forming galaxies. In the following section, we employ a similar approach on a sample of star-forming galaxies (as a comparison to the molecular outflow sample) in order to better understand the role star formation plays in the gamma-ray emission of our sample vs the outflow itself.

\subsubsection{Star-forming Galaxy Comparison Sample}\label{sec:SFG}
\begin{figure}
    \centering
    \includegraphics[width=3.5in]{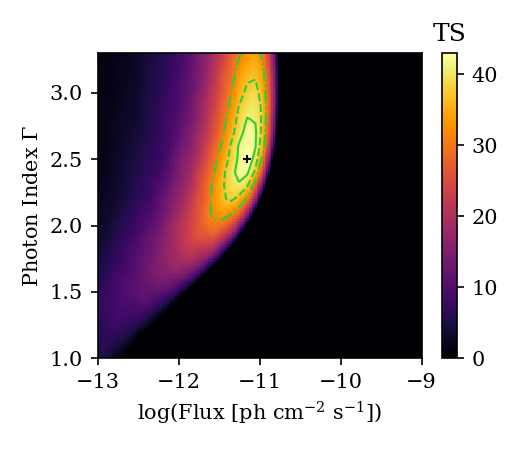} 
    \caption{Stacked TS profile for the sample of star-forming galaxies. Overlaid are the 1, 2, and $3\, \sigma$ contours for 2 degrees of freedom.}
    \label{fig:SFG_TS}
\end{figure}

\begin{figure}
    \centering
    \includegraphics[width=3.5in]{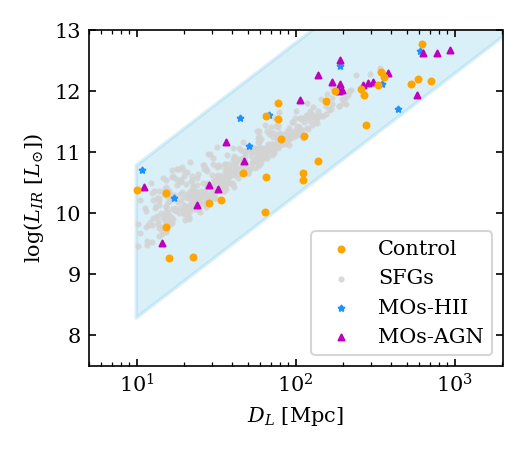}
    \includegraphics[width=3.5in]{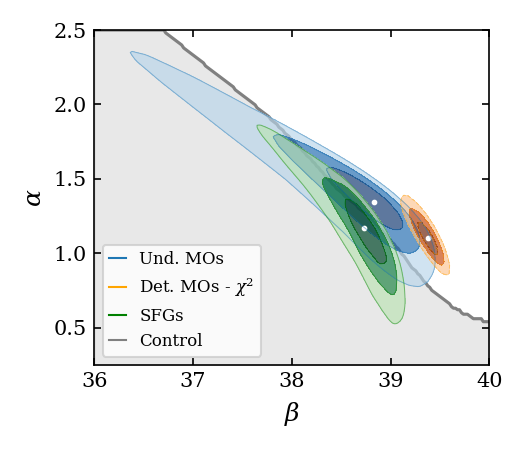}
    \caption{Left: Distribution in $L_{IR}-D_L$ space of our benchmark sample (blue stars for \ion{H}{ii} and magenta triangles for AGN galaxies), control sample (orange dots), and the SFG sample (grey dots). The shaded region shows our $L_{IR}-D_L$ selection criteria. Right: 1, 2, and $3\,\sigma$ $\alpha-\beta$ contours characterizing the $L_{\gamma}-L_{IR}$ correlation for the benchmark sample of undetected molecular outflows (blue), the detected outflows (orange), the SFG sample (green), and the 95\% upper limits for the Control sample (grey). Best-fit values are shown for the SFG and molecular outflow sample as white dots. }\label{fig:SampleComps_and_Contours}
\end{figure}

Many of the galaxies in our sample have significant star formation activity. The $L_{IR}-L_{\gamma}$ relation has been well established in previous studies of star-forming galaxies (SFGs, \citealt{ackermann2012}; \citetalias{ajello_sfgs}). In order to compare the gamma-ray signal in our sample with that of the previous works, we reanalyze a subset of the SFGs studied in \citetalias{ajello_sfgs} using our analysis pipeline as described in Section \ref{sec:analysis}. Beginning with the full sample used in that analysis, we employ the same catalog cross matching selection criteria as for our original sample, removing targets that are spatially coincident with 4FGL, BZCAT, and radio galaxy sources and removing any molecular outflows in our analysis. Furthermore, we limit the galaxies to those that are roughly compatible with the $L_{IR} - D_L$ distribution of our sample. Specifically, we keep only galaxies with $10\:\mathrm{Mpc} <D_L<1000\:\mathrm{Mpc}$ and $10^{6.5}\left(D_L/\mathrm{Mpc}\right)^2<L_{IR}/L_{\odot}<10^{8.8}\left(D_L/\mathrm{Mpc}\right)^2$. This ultimately leaves us with a sample of 515 star-forming galaxies as a comparison sample (hereafter referred to as the SFG sample). The IR luminosities and distances used for the SFG sample are taken from \citetalias{ajello_sfgs}. We briefly note that the \citetalias{ajello_sfgs} sample primarily consists of a subset of the IRAS RBGS \citep{rbgs}. The range of distances and IR luminosities used in the selection is shown as the shaded region in the left panel of Figure \ref{fig:SampleComps_and_Contours}. Also shown in this figure are the $L_{IR} - D_L$ values for the star-forming galaxies, our benchmark molecular outflow sample, and the control sample (see Section \ref{sec:alma_control}). {The analysis of the SFG sample yields a total TS of 42.9, with best fit index of $\Gamma= 2.5\lims{0.4}{0.2}$ and flux of $6.31\lims{1.6}{3.7}\times 10^{-12}$ ph cm$^{-2}$ s$^{-1}$. The TS profile for the SFG sample is shown in Figure \ref{fig:SFG_TS}.}

{This analysis yields a significant detection of gamma rays from SFGs consistent with previous work (\citetalias{ajello_sfgs}). However, it is worth considering why no signal is detected in the control sample while there is in the SFGs, since presumably the control sample would also produce some gamma rays from star formation. One factor is that the difference in the sizes of the control and the SFG samples affects the detection significance. We demonstrate this by repeatedly sampling a random subset of 30 galaxies from the SFG sample and computing the TS. We find that there is a roughly 30\% chance of obtaining a TS at or below the level of the control sample (TS $\lesssim 2$). In contrast, we find a $<0.5\%$ chance of obtaining a TS near the level of the benchmark sample (TS $>20$). Another important consideration is that the flux limit of the control sample is appreciably lower than the SFGs, as can be seen in the left panel of Figure \ref{fig:SampleComps_and_Contours} by comparing the lower edge of the blue shaded region with the distribution of SFGs (grey dots). The SFGs are based primarily on the flux-limited sample of the RBGS \citep{rbgs} and are therefore selected for bright IR galaxies, whereas the control sample is selected for galaxies with a flux limit based on the benchmark sample\footnote{We note that a more stringent flux limit (e.g. $\sim$3 times higher than that of Figure \ref{fig:SampleComps_and_Contours}) has negligible impact on the results for the control or benchmark samples.}. A final consideration is that in constructing the control sample we were careful to select galaxies that do not have molecular outflows. While we remove known molecular outflows from the SFG sample, it is possible that this sample contains galaxies hosting molecular outflows that have not been detected due to lack of direct observations and analysis. }

\begin{figure}
    \centering
    \includegraphics[width=6.8in]{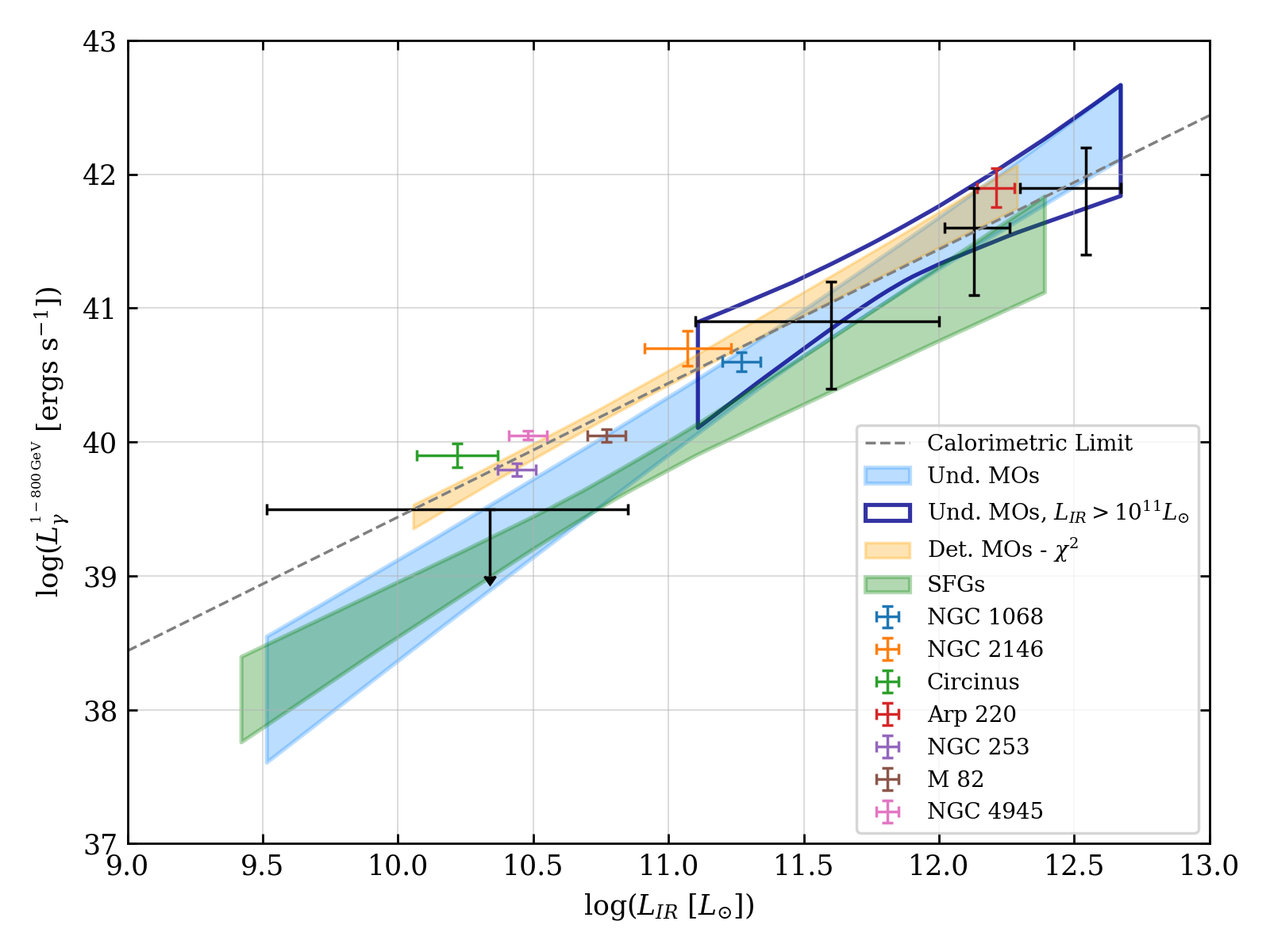}
    \caption{Plot of $L_\gamma$ vs  $L_{IR}$. Our molecular outflow sample is divided into 4 quantile bins based on $L_{IR}$, shown with the black crosses. {The abscissa is set at the mean $L_{IR}$ of each bin. For the lowest $L_{IR}$ bin we show the 95\% upper limit.} The grey dashed line is the calorimetric limit. Gamma-ray-detected galaxies known to host molecular outflows are plotted individually. We also show the $1\,\sigma$ band for {the undetected molecular outflows (blue), the SFG comparison sample (green), and the detected molecular outflows (orange). The dark blue contour shows the $1 \, \sigma$ band for the undetected molecular outflows with $L_{IR}>10^{11}L_{\odot}$}.}\label{fig:Lgamma_LIR}
\end{figure}

We also analyze a number of galaxies that have been detected in gamma rays and have also been observed to contain molecular outflows. These include the five gamma-ray-detected outflows from the \citetalias{fluetsch19} sample (NGC 253, NGC 1068, Circinus, M 82, and NGC 2146), as well as NGC 4945 \citep{lenain2010, ackermann2012, bolatto2021} and Arp 220 (\citealt{peng2016}; \citetalias{ajello_sfgs}; \citealt{arp220_SF_enhance}). 
Properties of these galaxies and their outflows are listed in Table \ref{tab:det_gals}. Each of these galaxies has been analyzed following the same procedures as the SFG and benchmark samples (see Section \ref{sec:data_selection}). Since the detected galaxies often have larger uncertainties in $\log (L_{IR})$ than $\log (L_{\gamma})$, we employ an orthogonal distance regression (ODR) method \citep{odr}. This method takes into account the two-dimensional uncertainties, which are not incorporated in the stacking method. The fit is evaluated using the ODR equivalent version of the $\chi^2$ metric. This yields best-fit values of $\alpha = 1.11\lims{0.08}{0.06}$ and $\beta = 39.37\lims{0.07}{0.05}$ with a reduced $\chi^2$ of $\chi^2/(d.o.f.) = 16.03/5=3.21$. While the reduced $\chi^2$ value does not indicate a good fit, this is likely due to the uncertainties in distance which have not been accounted for here, and which range in value from $\sim 5-15\%$, and some intrinsic scatter is to be expected based on previous results \citep{ajello_sfgs}. The resulting best-fit $\alpha-\beta$ values for the SFG sample, the undetected molecular outflows, and the individually-detected galaxies with outflows are provided in Table \ref{tab:alpha_beta_comp}. In this table, we also show the best-fit $\alpha-\beta$ values for the subsample of undetected molecular outflows with $L_{IR}>10^{11} L_{\odot}$.
In Figure \ref{fig:Lgamma_LIR}, we show the $1\sigma$ bands for the $L_{\gamma} - L_{IR}$ relation {for the undetected and detected molecular outflow samples as well as the sample of SFGs}. We also show the data points for the seven detected galaxies with molecular outflows and data points for the undetected molecular outflows stacked in bins of $L_{IR}$. Additionally, we include the calorimetric limit wherein the cosmic rays in the galaxy lose most of their energy to pion production of gamma rays, assuming a conversion efficiency of supernova energy to cosmic rays of $10\%$ (cf. \citealp{thompson2007, lacki2011,ackermann2012}). {The dark blue outlined region in Figure \ref{fig:Lgamma_LIR} shows the $1\, \sigma$ band for only undetected galaxies in our sample that have $L_{IR}>10^{11}L_{\odot}$. These targets dominate the signal and are consistent with both the calorimetric limit and the $1 \sigma$ band of the detected galaxies.}
\begin{table}[htbp]\label{tab:gal_params}
\centering

\begin{tabular}{lcHHcccccHHcccc}

\toprule
Name & Type & RA & Dec &$D_L$ & SFR & $\log L_{IR}$ & $\log L_{AGN}$ & $\alpha_{bol}$ & $R_{out}$ & $v_{out}$ & $\dot{M}_{out}$ & $\log P_k$ & $P_k/L_{AGN}$ &$q_{IR}$\\
 &  & [deg.]  & [deg.] &[Mpc] &  [$M_{\odot}$/yr] & [$L_{\odot}$] & [ergs/s] &  & [pc] & [km/s] & [$M_{\odot}$/yr] & [ergs/s]& & \\
\midrule
NGC 253 & \ion{H}{ii} & ra & dec &  3.30 & 2.8 & 10.44 & $\leq 40.66$ & $ \leq 4\times 10^{-4}$ & $R_{out}$ & 50 & 1.4 & 39.04 & $\geq 0.024$ & 3.01  \\

NGC 1068 & Sy2 & ra & dec & 13.10 & 16.8 & 11.27 & $43.94$ & $0.097$ & $R_{out}$ & 150 & 28.0 & 41.30 & 0.0023 & 2.32  \\

NGC 2146 & \ion{H}{ii} & ra & dec & 18.00 & 11.7 & 11.07 & $\leq 41.09$ & $\leq 3\times 10^{-4}$ & $R_{out}$ & 150 & 5.0 & 40.52 & $\geq 0.27$ & 2.83  \\

M 82 & \ion{H}{ii} & ra & dec &  3.70 & 5.9 & 10.77 & $\leq 41.54$ & $\leq 9\times 10^{-4}$ & $R_{out}$ & 100 & 4.0 & 40.09 & $\geq 0.036$ & 2.62  \\

Circinus & Sy2 & ra & dec &  4.21 & 0.7 & 10.22 & $43.57$ & $0.59$ & $R_{out}$ & 150 & 1.0 & 39.87 & $2\times 10^{-4}$ & 2.07  \\

NGC 4945 & Sy2 & ra & dec &  3.80 & 2.2 & 10.48 & 43.54 & 0.26 & $R_{out}$ & $\geq 240$ & 20.0 & 41.56 & 0.01 & 2.48  \\

Arp 220 & LINER & ra & dec & 79.90 & 134.6 & 12.21 & 45.08 & 0.17 & $R_{out}$ & $800$ & 100.0 & 43.31 & $0.017$ & 2.99  \\
\bottomrule
\end{tabular}\caption{Galaxy and outflow properties for the individually detected galaxies. Luminosity distances are taken from NED, and the infrared luminosities ($L_{IR}$) are computed from the IRAS fluxes. The SFR is computed using $L_{IR}$, the AGN contribution to the total bolometric luminosity, and the relation of \cite{sturm2011}. The AGN contribution to the total bolometric luminosity is given by $\alpha_{bol}=L_{AGN}/L_{bol}$. $P_k$ is the kinetic power of the outflow, defined as $P_k=0.5\dot{M}_{out}v^2_{out}$. The $\alpha_{bol}$, $L_{AGN}$, mass-loss rates, outflow velocities, and type classifications are taken from \citetalias{fluetsch19} for galaxies included in that study (i.e. all except NGC 4945 and Arp 220). $L_{IR}$ values are taken from \citetalias{ajello_sfgs} except for Circinus, which is taken from \citet{kornecki2020}. $q_{IR}$ is the ratio between the IR and 1.4 GHz radio fluxes (as defined in \citealt{helou1985, ivison2010, harrison2014}) with radio fluxes taken from NED. Logarithmic values for $L_{IR}$, $L_{AGN}$, and $P_k$ are base 10 (i.e. $\log_{10}$). The outflow mass-loss rate and velocity for NGC 4945 are taken from \citet{bolatto2021}, the AGN luminosity is from \citet{marconi2000}, and the classification is from \citet{baumgartner}. For Arp 220, the outflow mass-loss rate and velocity are from \citet{barcosMunoz2018}, and the classification, $\alpha_{bol}$, and AGN luminosity are from \citet{nardini2010}.} \label{tab:det_gals}
\end{table}

The right panel of Figure \ref{fig:SampleComps_and_Contours} shows the 1, 2, and 3 $\sigma$ contours for the SFG sample, the undetected molecular outflow sample, and the detected outflow sample. For the control sample, we show the 95\% upper limit on the $\beta$ parameter computed using the ``delta-log-likelihood'' method by finding $2\Delta\log L = -2.71$ \revision{for each $\alpha$ value} (see e.g. \citealt{2014PhRvD..89d2001A,2015PhRvL.115w1301A,magic2016}). The contours of the undetected molecular outflow sample are consistent with those of the SFG comparison sample, although the undetected molecular outflow sample has a greater {best-fit} $\alpha$ dependence. The results for the control sample are also compatible with the undetected molecular outflow and SFG contours, in particular noting that the SFG contours are essentially fully enclosed up to the 3 $\sigma$ level within the control sample upper limit region. {The blue molecular outflow contours are largely contained within the grey control region; however, the best-fit point and most of the 1 $\sigma$ contour are outside of this region, indicating that there may be some level of gamma-ray emission due to the presence of the outflow.} Furthermore, the complete compatibility of the SFG contours with the control sample upper limits suggests that the SFG sample is likely comprised of galaxies that lack prominent molecular outflows.

\begin{table*}[t]
\centering
\def\arraystretch{1.5}
\setlength{\tabcolsep}{12pt}
\begin{tabular}{ccccc}
	\hline\hline
 &$\Gamma_{\gamma}$&$\alpha$ &$\beta$ & TS$_{max}(\alpha,\beta)$  \\
\hline
SFGs&$2.5\lims{0.4}{0.2}$& $1.15\lims{0.09}{0.12}$ & $38.73\lims{0.10}{0.10}$ & 52.5\\
Undetected MOs& $2.0\lims{0.3}{0.3}$ &$1.36\lims{0.08}{0.12}$ & $38.71\lims{0.16}{0.20}$ & 25.9\\
Und. MOs ($L_{IR} > 10^{11}L_{\odot}$)& $2.0\lims{0.3}{0.2}$ &$1.18\lims{0.09}{0.13}$ & $39.20\lims{0.16}{0.25}$ & 20.2\\


Detected MOs &$2.33\lims{0.07}{0.06}$ &$1.11\lims{0.08}{0.06}$&$39.37\lims{0.07}{0.05}$&$\chi_{red}^2 = 3.21$\\
\hline\hline
\end{tabular}
\caption{ Best-fit $\alpha-\beta$ values and corresponding TS (in the $\alpha-\beta$ plane) for the different samples, following the $L_{\gamma}-L_{IR}$ relation of Equation \ref{eq:Lgamma_LIR}. We also show the best-fit photon index $(\Gamma_{\gamma})$ from the flux-index stack for each sample. In the bottom row, we show the best-fit parameters for the gamma-ray-detected sample obtained using the $\chi^2$ metric.
}\label{tab:alpha_beta_comp}
\end{table*}

In many cases, particularly for highly star-forming galaxies, the IR luminosity of a galaxy can be considered a direct proxy for the SFR in a galaxy \citep{kennicutt1998,  bell03,kennicuttEvans2012}. However, in some cases the central AGN can be a significant contributor to the total IR luminosity. We account for this by using the estimated AGN contribution to the total luminosity as provided in \citetalias{fluetsch19} by the parameter $\alpha_{bol}=L_{AGN}/L_{bol}$. Making use of the $L_{IR}$ is helpful in comparing our results to previous studies of gamma rays from star-forming galaxies; however, by accounting for the AGN contribution, we can obtain a more realistic estimate of the SFR and a better understanding of the role of star formation in our molecular outflow sample. We adopt the AGN-corrected star-formation rate from \citet{sturm2011}, which is of the form 
\begin{equation}
    SFR\left(\mathrm{M}_{\odot} \, \mathrm{yr}^{-1}\right) = (1-\alpha)\times 10^{-10} L_{IR}.
\end{equation}
Using the relation
\begin{equation}
        \log_{10} \left(\frac{L_{\gamma}}{\mathrm{ergs/s}}\right) = \beta + \alpha \log_{10} \left(\frac{\mathrm{SFR}}{\mathrm{M}_{\odot} \, \mathrm{yr}^{-1}}\right),
\end{equation}
we find $\alpha= 1.43\lims{0.09}{0.11}$ and $\beta=38.71\lims{0.17}{0.23}$ with a TS value of TS = 28.8. We note that the overall $\alpha-\beta$ dependence is compatible with the $L_{\gamma}-L_{IR}$ relation, though we see a slight improvement in the TS when comparing the $L_{\gamma}$ with the SFR when accounting for the AGN contribution.

\section{Conclusions}\label{sec:conclusion}

In this work, we have performed a stacked gamma-ray analysis of a sample of nearby galaxies known to host molecular outflows. The results of this analysis provide evidence of gamma-ray emission from this population, particularly in contrast to the lack of signal seen in the control sample of galaxies without molecular outflows. In the analysis of only those targets in our sample that are not individually resolved, we find a detection of gamma-ray emission at a significance of $4.4\, \sigma$ with an average photon flux for the sample of $1.3\lims{0.7}{0.6}\times 10^{-11}$ ph cm$^{-2}$ s$^{-1}$ with photon index $\Gamma=2.0\lims{0.3}{0.2}$ {in the $1-800$ GeV range}. The bulk of this signal can be attributed to {\ion{H}{ii} galaxies and other AGN galaxies consistent with an “energy-conserving” driving mechanism as indicated by high kinetic power to AGN luminosity ratios.}  

We do not find strong evidence for direct scaling of the gamma-ray luminosity with properties intrinsic to the outflows themselves (e.g. outflow mass rate and kinetic power). In other words, there is no evidence that the outflow is directly accelerating cosmic rays.
Rather, the most prominent scaling with the gamma-ray luminosity is with properties of the host galaxy – namely, the infrared luminosity (and in turn the related SFR). 
{In comparison with the SFG sample, galaxies hosting molecular outflows tend to exhibit somewhat different $L_{\gamma}-L_{IR}$ properties. In the case of the detected outflow-hosting galaxies, the distinction is highly pronounced as this sample occupies an entirely different region of the $L_{\gamma}-L_{IR}$ plane. For the sample of undetected galaxies with molecular outflows, the distinction is not as stark. While the SFGs and undetected galaxies hosting outflows are mostly compatible, there is deviation in the $L_{\gamma}-L_{IR}$ scaling relation parameter $\alpha$ that can be seen in both the $\alpha-\beta$ contour plot (Figure \ref{fig:SampleComps_and_Contours}) and the $L_{\gamma}-L_{IR}$ relations (Figure \ref{fig:Lgamma_LIR}). In fact, as can be seen in Figure \ref{fig:Lgamma_LIR}, the differences in these parameters result in compatibility between the undetected galaxies hosting outflows and several of the individually detected galaxies (especially M 82, NGC 1068, and Arp 220), whereas these are still outliers from the SFG band. Figure \ref{fig:Lgamma_LIR} also shows that the sample of gamma-ray detected galaxies with an outflow are on average near perfect calorimeters, in contrast to the full sample of undetected molecular outflows or SFGs.} Although, as demonstrated in Table \ref{tab:alpha_beta_comp} and Figure \ref{fig:Lgamma_LIR}, the subsample of undetected galaxies with molecular outflows classified as LIRGs or ULIRGs (i.e. $L_{\odot}>10^{11}$ or $L_{\odot}>10^{12}$) are also compatible with the calorimetric limit. Additionally, the galaxies in our sample have radio-infrared ratios (as indicated by $q_{IR}$) compatible with typical star-forming galaxies. This suggests that the galaxies in our sample are not accelerating cosmic ray electrons to a greater extent than other star-forming galaxies and that any cosmic-ray protons present are efficiently converted into gamma rays, consistent with the observed calorimetry.

In a number of galaxies, recent observational evidence has been found for star formation triggered within the outflow itself \citep{ SFinOutflow, gallagher2019,arp220_SF_enhance}. The triggering of star formation in outflows is a consequence of higher density regions caused by compression of the cold gas swept up in the expanding shock. Within these local density enhancements, the rate of proton-proton interactions {could potentially increase}, which may in turn  produce additional gamma rays. Thus, it is possible that the molecular outflow enhances a galaxy's gamma-ray emission in these regions.  

In addition, the observed calorimetry of the gamma-ray-detected sample and the sample of undetected high-$L_{IR}$ galaxies suggests that galaxies hosting molecular outflows may be bright sources of high-energy neutrinos, as evidenced by the marginal detection of NGC 1068 by IceCube \citep{aartsen2020}. Indeed, starburst galaxies are expected to accelerate protons and produce neutrinos up to high TeV and PeV energies \citep{tamborra2014,yoastHull2015,peretti2020, ha2021}, and the presence of molecular outflows may play a meaningful role in the neutrino production given the {near} calorimetric nature of the {detected} population.

Increasingly, it appears that molecular outflows are a common feature in galaxies, and in particular molecular outflows seem to be highly common in ULIRGs \citep{ chen2010, veilleux2013, perSant2018,puma2}. For instance, nearly all of the ULIRGs in the GOALS \citep{goals} and IRAS RBGS \citep{rbgs} catalogs have detected or tentative evidence of a molecular outflow. It may be the case that molecular outflows are a commonality in gamma-ray-emitting SFGs, particularly those detected in gamma rays at greater distances. In fact, several of the SFGs detected by {\Fermi} also have well observed molecular outflows (e.g. NGC 253, NGC 1068, NGC 2146, NGC 4945, Circinus, M82, Arp 220). While our analysis suggests the outflow itself may not be responsible for the direct acceleration of cosmic rays, it may enable an environment favorable to efficient conversion of cosmic rays to gamma-ray emission, for example by way of enhanced star formation in the outflow.

Future studies may be able to probe molecular outflows and their gamma-ray emission more carefully through increased sample sizes and more in depth studies of their outflow properties. Ongoing and planned surveys as well as dedicated studies continue to discover new evidence of molecular outflows in both local and more distant galaxies \citep{ lutz2020,ngc1482, may2020, stuber2021, bolatto2021}. Additionally, studies of more distant molecular outflows (see e.g. \citealp{stuber2021}) and explorations of trends between their {gamma-ray and infrared luminosities} in conjunction with SFGs and ULIRGs at these larger redshifts can also perhaps clarify the relationship between the molecular outflows, star formation, and the gamma-ray emission. The presence of gamma-ray emission in the molecular outflows studied in this paper sets up an intriguing path for determining to what extent the molecular outflow itself plays a role in the production of gamma rays, and the results here can aid in the development of theoretical modeling to disentangle contributions from star formation and from the molecular outflow. 

\begin{acknowledgments}
The authors acknowledge support from NASA grant 80NSSC21K1915. Clemson University is acknowledged for generous allotment of compute time on Palmetto cluster. 

The \textit{Fermi} LAT Collaboration acknowledges generous ongoing support
from a number of agencies and institutes that have supported both the
development and the operation of the LAT as well as scientific data analysis.
These include the National Aeronautics and Space Administration and the
Department of Energy in the United States, the Commissariat \`a l'Energie Atomique
and the Centre National de la Recherche Scientifique / Institut National de Physique
Nucl\'eaire et de Physique des Particules in France, the Agenzia Spaziale Italiana
and the Istituto Nazionale di Fisica Nucleare in Italy, the Ministry of Education,
Culture, Sports, Science and Technology (MEXT), High Energy Accelerator Research
Organization (KEK) and Japan Aerospace Exploration Agency (JAXA) in Japan, and
the K.~A.~Wallenberg Foundation, the Swedish Research Council and the
Swedish National Space Board in Sweden.
 
Additional support for science analysis during the operations phase is gratefully
acknowledged from the Istituto Nazionale di Astrofisica in Italy and the Centre
National d'\'Etudes Spatiales in France. This work performed in part under DOE
Contract DE-AC02-76SF00515.

This research has made use of the NASA/IPAC Extragalactic Database (NED) which is operated by the Jet Propulsion Laboratory,
California Institute of Technology, under contract with the National Aeronautics and Space Administration..
\end{acknowledgments}

%

\vspace{5mm}
\facilities{\Fermi-LAT \citep{atwood2009}}


\software{\texttt{astropy} \citep{astropy},  
          \texttt{Fermipy} \citep{fermipy}}






\bibliography{ref}
\bibliographystyle{aasjournal}



\end{document}